# Magnetic-field enhanced high-thermoelectric performance in topological Dirac semimetal $Cd_3As_2$ crystal

H. H. Wang[a,†], X. G. Luo[a,†], W. W. Chen[a], N. Z. Wang[a], B. Lei[a], F. B. Meng[a], C. Shang[a], L. K. Ma[a], T. Wu[a], X. Dai[b], Z. F. Wang[a], X. H. Chen[a,c,d,*]

[a] Hefei National Laboratory for Physical Science at Microscale and Department of Physics, University of Science and Technology of China, and Key Laboratory of Strongly-coupled Quantum Matter Physics, Chinese Academy of Sciences, Hefei, Anhui 230026, China.

[b] Physics Department, Hong Kong University of Science and Technology, Clear Water Bay, Kowloon, Hong Kong

[c] High Magnetic Field Laboratory of the Chinese Academy of Sciences, Hefei, Anhui 230031, China.

[d] Collaborative Innovation Center of Advanced Microstructures, Nanjing University, Nanjing 210093, China.

*Correspondence to: chenxh@ustc.edu.cn.
† Equal contributions.

**Thermoelectric materials can be used to convert heat to electric power through the Seebeck effect. We study magneto-thermoelectric figure of merit (*ZT*) in three-dimensional Dirac semimetal $Cd_3As_2$ crystal. It is found that enhancement of power factor and reduction of thermal conductivity can be realized at the same time through magnetic field although magnetoresistivity is greatly increased. *ZT* can be highly enhanced from 0.17 to 1.1 by more than six times around 350 K under a perpendicular magnetic field of 7 Tesla. The huge enhancement of *ZT* by magnetic field arises from the linear Dirac band with large Fermi velocity and the large electric thermal conductivity in $Cd_3As_2$. Our work paves a new way to greatly enhance the thermoelectric performance in the quantum topological materials.**

## 1. Introduction

Thermoelectric (TE) materials play a significant role in the energy conversion. The efficiency of such conversion depends on a dimensionless figure of merit (*ZT* =$(S^2/\rho)(T/\kappa)$), where *S* is the Seebeck coefficient, $\rho$ is the electrical resistivity, *T* is the absolute temperature and $\kappa$ is the thermal conductivity. The research on the TE materials for seeking high *ZT* started decades ago. How to improve the conversion efficiency is a hot topic due to the increase of energy demand and environment-friend energy. For a good TE material, high Seebeck coefficient *S* and low electrical resistivity $\rho$, and low thermal conductivity $\kappa$ are required to be satisfied at the same time. However, *S*, $\rho$ and $\kappa^e$ (electronic contribution to $\kappa$) are usually highly coupled, so that *ZT* can be only tuned within some limitations. Though there is no theoretical limit for value of the *ZT*, the maximum *ZT* of ~1 at room temperature is observed in $Bi_2Te_3$, and is far from the extensive utilization of thermoelectric technology [1]. Usually, there are two routes to be used to enhance *ZT* with increasing the power factor $S^2/\rho$ or decreasing the lattice thermal conductivity by so-called phonon engineering. To substantially reduce the lattice thermal conductivity, the phonon engineering extensively increases the phonon scattering through defect engineering [2,3], nanostructuring [4] (such as thin film [5], superlattice [6] and quantum dots [7]), or making the crystal structure complicated [1]. The another approach, enhancing power factor, has been realized by band engineering [2,8,9], inducing resonant level [10], ionized impurity scattering [11], modulation doping [12], and increasing carrier mobility [3].Very recently, Zhao *et al.* reported that magnetic-nanoparticles embedded in the thermoelectric matrix can tune the independent phonon- and electron-transport properties at the same time to efficiently enhance *ZT* [13].

It is well known that some of the good thermoelectric materials are also typical topological materials: $Bi_2Te_3$ and $Sb_2Te_3$ with topological surface states at which the conduction and valence bands touch each other at the Dirac-Cone with linear energy dispersion in all momentum directions, and the carriers have high mobility and small effective mass [14,15]. As an analogue, the 3D topological Dirac semimetal $Cd_3As_2$ shares similar electronic structure. The topological Dirac semimetal of $Cd_3As_2$ is fourfold degenerate three dimensional (3D) Dirac materials in which electrons of the bulk state have high mobility and obey linear energy dispersion, while the surface state is topology protected Fermi arc [16]. It has been reported that the magnetic field has a strong effect on the electric and thermal transport in topological Dirac and Weyl semimetals [17]. Small perturbations like strain, magnetic field and chemical potential may tune the 3D bulk Dirac states therein [18-20], even pressure or point-contact-induced superconductivity. A non-saturation large linear magnetoresistance is observed in $Cd_3As_2$ [21,22], such behavior is common feature in the topological materials with the linear energy dispersion in all momentum directions [21-23]. $Cd_3As_2$ is expected to be great potential for high thermoelectric performance because the ultrahigh electron mobility ($\mu \sim 10^4$-$10^6$ cm$^2$ V$^{-1}$ s$^{-1}$) [21] may lead to the possible large power factor $S^2/\rho \approx \mu(m^*/m_e)$, where $m^*$ and $m_e$ is the energy-band electron effective mass and the free electron mass, respectively [8]. However, the experimentally achieved maximum *ZT* for $Cd_3As_2$ is only about 0.15 [24]. An anomalous thermal conductivity had been reported in $Cd_3As_2$ five decades ago [25]. The contribution from lattice to the thermal conductivity is quite small in $Cd_3As_2$ due to anion vacancy, and even much less than the contribution from electron to the thermal conductivity at room temperature [25]. Therefore, this provides possibility to dramatically enhance *ZT* by magnetic field to suppress the electric thermal conductivity in $Cd_3As_2$. It is quite different from the usual way to reduce the lattice thermal conductivity by the phonon engineering. In this report, we report magneto-thermoelectric figure of merit in 3D topological Dirac semimetal $Cd_3As_2$ crystal. We systematically measured temperature dependence of resistivity, thermopower and thermal conductivity under the perpendicular magnetic field in

$Cd_3As_2$ crystal. To study the anisotropy of *ZT*, the crystals with [100] and [112] orientation from the same batch were used for the measurements under magnetic fields applied along [100] or [112] direction and perpendicular to electric or heat current, respectively. It is unexpected that the *ZT* is remarkably enhanced by a perpendicular magnetic field.

## 2. Experimental

The $Cd_3As_2$ single crystals were grown by vapor transport method, as described in our previous report [26]. Polycrystalline $Cd_3As_2$ was in prior synthesized by heating the stoichiometric mixtures of Cd and As powder in vacuum at 800 °C for 6 hours. The $Cd_3As_2$ powder was loaded into a quartz tube for the growth of single crystals. The quartz tube was then evacuated, sealed and placed into a two-zone horizontal furnace. To optimize the nuclei formation, the temperature of the "cold" end (empty) and the "hot" end (containing $Cd_3As_2$ powders) of the tube were first raised to 520°C and 575°C, respectively, held for 6 h, then cooled to 510°C and 560 °C in half an hour, respectively, and then kept there for 7-10 days. The sample with the largest dimensions of 7 mm×2 mm×2 mm can be achieved.

Single crystal platelets were characterized by using x-ray diffraction (XRD), performed on a SmartLab-9 diffractometer (*Rikagu Inc.*) from 10° to 70° with a scanning rate of 4° per minute. The achieved large crystals were usually oriented naturally along [100] or [112] directions (see Fig.S1 in the Supplementary Materials). Crystals with both of [100] and [112] orientations were selected for thermoelectric measurements. Crystals were carefully polished to regular rectangular shape by keeping the flat large surface with [100] or [112] orientations unchanged before the transport measurements. The transport measurements including resistivity, thermopower and thermal conductivity were carried out by using PPMS-14T system (*Quantum Design*). Longitudinal and Hall resistivity were measured by standard six-probe method. During the measurements, magnetic field is always applied along the crystal with [100] or [112] orientations, and electric or heat currents ($I$ and $\nabla T$) are perpendicular to the crystal orientation direction. Longitudinal resistivity in magnetic fields were measured by using the two same-side probes and then symmetrized between the positive and negative fields ($\rho(B)=[\rho(+B)+\rho(-B)]/2$) to cancel the picked-up Hall signal. Hall resistivity ($\rho_{xy}$) was measured by using the two opposite-side probes calculated through the form of $\rho_{xy}(H)=[\rho(+B)-\rho(-B)]/2$. Resistivity, thermopower and thermal conductivity in high temperatures range from 350 K to 500 K and under magnetic field up to 9 Tesla were measured by a home-built system in a bottom-loading superconducting magnet system (*American Magnetic Inc.*) with a water-cooling-protected Ta-resistance furnace inserted to provide high temperature. High-temperature thermopower and thermal conductivity were collected by adopting the longitudinal steady-state thermal flow method in high vacuum (~$1\times10^{-4}$ Pa), with E-type chromel-constantan thermocouples of 10 μm diameter from Omega measuring the temperature differential and phosphor-bronze wires of 10 μm diameter as electric connection to measure thermopower. Samples were mounted on a copper heat-sink with [100] or [112] axis parallel to the magnetic field but perpendicular to $I$ or $\nabla T$. Heat flow for establishing the temperature gradient was supplied by a resistor chip ($R$=1820Ω) heater, which is mounted on the hot side of the sample and thermally insulated from the copper heat-sink by a 25 cm long and 18 μm-diameter constantan wires. Small temperature gradient (~1 K) was used for thermal measurements to minimize the proportion of heat loss relative to total heat flow supplied by the heater. The radiation loss was estimated as $P_{rad}=\sigma_T \times (A_{heater}+A_{sample}/2) \times (T_{hot}^4-T_{cold}^4)$, where $\sigma_T=5.67\times10^{-8}$ Wm$^{-2}$K$^{-4}$ is the Stefan-Boltzmann constant, $A_{heater}$ and $A_{sample}$ are the surface area of the resistor chip heater and the surface area of the sample, and $T_{hot}$ ($T_{cold}$) is the temperature of hot (cold) side of the sample. The actual heat flowing through the sample was then estimated by subtracting the radiation loss from the total heat flow supplied by the heater.

## 3. Results and Discussion

Figure 1 shows temperature dependence of resistivity in the temperature range from 2 to 500 K under magnetic fields ($B$) up to 14 Tesla applied along [100] or [112] direction and perpendicular to electric current, respectively. The residual resistivity is estimated by fitting the data below 50 K with the formula of $\rho(T)=\rho_0+AT^2$ to be $\rho_0 =$ 0.489 and 0.587 μΩ m for the crystals with [100] and [112] orientation, respectively. The residual resistance ratios (RRR=$\rho$(300K)/$\rho_0$) are 6.0 and 5.6 for the crystals with [100] and [112] orientation, respectively. As shown in Fig.1, large positive magnetoresistivity is observed for the magnetic field applied along [100] or [112] direction, and resistivity monotonically increases with increasing magnetic field, being consistent with the isothermal magnetoresistivity data at 300 K as shown in Fig.S2 of Supplementary Materials. The magnetoresistivity (MR=[$\rho(B)$-$\rho$(0T)]/$\rho$(0T)) is as high as 2600% at 3 K and 680% at 300 K for the magnetic field applied along [100] direction, and 2400% at 3 K and 260% at 300 K for the magnetic field applied along [112] direction. It indicates that the MR for the magnetic field applied along [112] direction is much less than that for magnetic field applied along [100] direction at room temperature although their MR is nearly the same at low temperature. As shown in Fig.1a, the crystal always shows metallic behavior in the whole temperature range for the case of the magnetic field applied along [100] direction. However, the crystal shows the metallic behavior only under magnetic field less than 4.5 Tesla, and $\rho(T)$ changes the slope from positive to negative at certain temperature with decreasing temperature above 4.5 Tesla for the case of the magnetic field applied along [112] direction. Such transition from metal to semiconductor arises from the quantum limit under large perpendicular magnetic field.

Figure 2 displays Seebeck coefficient as a function of temperature under various magnetic fields applied along [100] or [112] orientation and perpendicular to temperature gradient, respectively. Without the applied magnetic field, $S$ is negative with values of -60 and -63 µV K$^{-1}$ at 300 K for the crystals with [100] and [112] orientation, respectively, and shows a temperature-linear dependence. Such behavior can be interpreted by the Seebeck coefficient obtained from the semi-classical Boltzmann equation ($S = -\frac{5}{24}\frac{\gamma^2}{v_f^6 \hbar^6}\frac{k_B}{e} k_B T \varepsilon_f$, with $k_B$ the Boltzmann constant, $e$ the electron charge, $v_f$ the Fermi velocity, $\varepsilon_f$ the Fermi energy and $\gamma$ a short-range disorder related parameter) for a 3D Dirac system [27]. $S$ pronouncedly increases in magnitude with increasing the magnetic field, and still shows linear $T$-dependence under magnetic fields. As shown in Fig.S3a and S3c in Supplementary Materials, the isothermal $S$ increases with increasing the perpendicular magnetic field, then saturates. The $B$-dependent isothermal $S$ obeys the formula $\Delta S = S(B) - S_0 = (S_\infty - S_0)\frac{\mu_T^2 B^2}{1+\mu_T^2 B^2}$, accordingly, it can be rewritten in the form of $\frac{B^2}{\Delta S} = \frac{1}{\Delta S_\infty}\left(\frac{1}{\mu_T^2} + B^2\right)$, where $S_0$ is the zero-field thermopower, $S_\infty$ is the limitation thermopower at infinite magnetic field, and $\Delta S_\infty = S_\infty - S_0$ [28]. As shown in Fig.S3b and S3d in Supplementary Materials, $\frac{B^2}{\Delta S}$ against $B^2$ shows a well-defined linear dependence, suggesting that the experimental results are in good agreement with the above formula. It should be addressed that the temperature-linear thermopower is kept up to 500 K under magnetic fields as shown in Fig.2b and 2d.

Figure 3 shows temperature dependence of the thermal conductivity $\kappa$ under different magnetic fields applied along the [100] or [112] direction and perpendicular to temperature gradient, respectively. The zero-field thermal conductivity shows a rapid increase with decreasing temperature below 50 K and a nearly $T$-independent behavior in the temperature range from 50 to 350 K, being consistent with the previous reports [29,30]. $\kappa$ at room temperature is about 3.3 and 3.0 W m$^{-1}$ K$^{-1}$ for the crystals with [100] and [112] orientation, respectively. As shown in Fig.3, the thermal conductivity is strongly suppressed when magnetic field is applied along [100] or [112] direction and perpendicular to the thermal gradient. $\kappa$ at the temperatures below 80K is very sensitive to the magnetic field, and is completely suppressed at low field and remains almost unchanged under the magnetic field higher than 2 Tesla. $\kappa$ at the temperatures above 80 K decreases gradually with further increasing the magnetic field, and reaches a saturation at about 7 Tesla. The saturation indicates a complete suppression of electronic contribution in the total $\kappa$. Such change of $\kappa$ with magnetic field is also confirmed by magnetic field dependent isothermal thermal conductivity as shown in Fig.S4 of Supplementary Materials, being consistent with the previous reports [30]. As shown in Fig.3a and 3c, $\kappa$ at 350 K is suppressed from 3.3 W m$^{-1}$ K$^{-1}$ without magnetic field to 0.71 W m$^{-1}$ K$^{-1}$ under magnetic field of 14 Tesla for the crystal with [100] orientation, and from 3.0 to 0.83 W m$^{-1}$ K$^{-1}$ for the crystal with [112] orientation, respectively. However, the thermal conductivity starts to increase with increasing the temperature above 350 K under the same magnetic field as shown in Fig.3b and 3d. It suggests that the lattice thermal conductivity increases with increasing the temperature above 350 K and gradually becomes the dominant contribution to the thermal conductivity at the high temperature.

According to the expression $ZT = (S^2/\rho)(T/\kappa)$, the increase of thermopower and the decrease of thermal conductivity by the perpendicular magnetic field lead to enhancement of *ZT*, while the positive magnetoresistivity hinders such enhancement. On the basis of *B*-dependent $\rho$, *S*, and $\kappa$ as shown in Fig.1-3, we calculated the corresponding dimensionless figure of merit. *ZT* at 350 K is 0.15 and 0.17 for the crystals with [100] and [112] orientation, respectively. As magnetic field is applied along [100] or [112] direction and perpendicular to the temperature gradient, *ZT* increases first with increasing magnetic field, and then decreases with further increasing magnetic field. The optimized magnetic field for obtaining the maximum *ZT* can be qualitatively explained in the following way. In our experiment, both thermal conductivity and thermopower saturate at high magnetic fields. However, the magnetoresistivity doesn't follow this rule, which monotonically increases with increasing magnetic field. Therefore, after passing a critical magnetic field, $\kappa$ and *S* saturate, but $\rho$ still increases. Therefore, *ZT* reaches a maximum value at the critical magnetic field. As observed in Fig.2c and 3c, $\kappa$ and *S* saturate under ~7 Tesla for the crystal with [112] orientation, and its *ZT* reaches the maximum value of 1.2, which is the largest *ZT* reported for all thermoelectric materials around 375 K so far, under ~7 Tesla and at 375 K as shown in Fig. 4d.

It should be addressed that there exists a large anisotropic enhancement of *ZT* by the perpendicular magnetic field for the crystals with [100] and [112] orientation. Without magnetic field, *ZT* shows nearly the same value and similar *T*-dependence for the crystals with [100] and [112] orientations. However, the *ZT* at 350 K is enhanced by the perpendicular magnetic field of 7 Tesla from 0.15 to 0.37 (by a factor of 2.47) for the crystals with [100] orientation, while from 0.17 to 1.1 (by a factor of 6.47) for the crystals with [112] orientation. Such large anisotropic *ZT* enhanced by magnetic field mainly arises from the anisotropic magnetoresistivity due to achievement of the quantum limit corresponding to different magnetic field as shown in Fig.1.

Although the positive magnetoresistivity is very large, a clear enhancement of power factor $S^2/\rho$ is observed in the high temperatures as shown in Fig.S5 because the Seebeck coefficient is drastically increased by the perpendicular magnetic field. The power factor is nearly enhanced by 2 at 350 K under magnetic fields of ~3 Tesla. At the same time, the thermal conductivity is remarkably reduced at the low fields below 3 Tesla, and saturates in the high fields larger than 4 Tesla as shown in Fig.S4. These results suggest that we can realize the enhancement of power factor and the reduction of thermal conductivity at the same time through magnetic field in the high temperatures. To understand the huge enhancement of *ZT* in the 3D topological Dirac semimetal $Cd_3As_2$, we also measured magneto-thermoelectric figure of merit in the P-doped $Cd_3(As_{0.54}P_{0.46})_2$ crystal with [100] orientation. As shown in Fig.S7, the thermal conductivity of the $Cd_3(As_{0.54}P_{0.46})_2$ crystal becomes small, while the resistivity and the thermopower greatly increases relative to the $Cd_3As_2$ crystal. In contrast to the $Cd_3As_2$ crystal, the magnetic field has an ignorable effect on thermopower and thermal conductivity for the $Cd_3(As_{0.54}P_{0.46})_2$ crystal, so that the *ZT* monotonously decreases with increasing magnetic field. Such different magnetic field-dependence of *ZT* might arise from the different levels of disorder scattering or Fermi energy level far from the Dirac points.

Compared to the conventional thermoelectric materials, the perpendicular magnetic field has a giant effect on the thermoelectric properties for the 3D topological Dirac semimetal $Cd_3As_2$. In order to get a deep physical understanding about these experimental observations, we have done a theoretical calculation based on the standard semi-classical Boltzmann equation (see Supplementary Materials). Under the magnetic field, to the leading order, a renormalized electric thermal conductivity can be written as $\kappa^e(B) = \kappa^e(0)(1-\omega_c^2\tau^2)$, where $\kappa^e(0)$ is electronic thermal conductivity without magnetic field, $\omega_c = eBv_f^2/\varepsilon_f$ is cyclotron frequency, $\varepsilon_f$ is Fermi energy, and $\tau$ is scattering time. Clearly, the large Fermi-velocity in Cd$_3$As$_2$ greatly reduces its electric thermal conductivity with increasing magnetic field. Since the total thermal conductivity has contributions from both electronic and lattice part, and previous theoretical work has already demonstrated that the electronic part is dominant in the thermal conductivity of Cd$_3$As$_2$ at high temperature [31]. Therefore, a giant reduction in thermal conductivity by the perpendicular high magnetic field is expected at high temperature, so that only lattice thermal conductivity is left. Extracting the lattice thermal conductivity from the experimental data, we can get a very good fitting for the total thermal conductivity (see Fig.S8a in Supplementary Materials). The value and the overall tendency for the thermal conductivity are in good agreement between theory and experiment. Recently, the violation of Wiedemann-Franz law has been experimentally reported in graphene, and is ascribed to the additional contribution to thermal transport by "neutral plasmon" excitation in Dirac fermion system [32]. One might expect a different magnetic field dependence for this additional thermal conductivity term, which is beyond the standard calculations in our report. In addition, the good agreement between the standard calculations and experiment suggests that the additional thermal conductivity by "neutral plasmon" excitation should be quite small.

Similar discussions can also be used for the thermopower. Under the perpendicular magnetic field, the thermopower can be calculated numerically, and depends on both cyclotron frequency and scattering time. Using the same parameters for fitting the thermal conductivity, the experimental data can be well fitted by the theoretical thermopower (see Fig.S8b in Supplementary Materials).

To increase the *ZT*, previous works mainly focus on the phonon-engineering to reduce the lattice thermal conductivity. In this report, as an alternative way, we introduce a novel way of electron-engineering to reduce the electric thermal conductivity by the perpendicular magnetic field, achieving a large *ZT* as high as 1.19 around 375 K. Traditionally, the magnetic field has a negligible effect on the thermal conductivity. This phenomenon can be understood in two aspects. Firstly, it is well known that electron will be localized by a perpendicular magnetic field due to the formation of Landau levels, so the electric conductivity decreases with increasing magnetic field. When the magnetic field is getting strong, the system is quickly approaching the quantum limit, so that the electric transport perpendicular to the magnetic field will be mainly contributed by inter-Landau level hopping. Such hopping will be quickly suppressed by the increment of cyclotron frequency ( $\omega_c = eBv_f^2 / \varepsilon_f$ ), which is proportional to the Fermi-velocity and magnetic field strength. Without violation of the Wiedemann-Franz law, the electric thermal conductivity and electric conductivity should have the same dependence on the magnetic field [17]. One notes that the conventional thermoelectric materials can be described by a parabolic band.

Under the magnetic field, its electric thermal conductivity is expressed as $\kappa^e(B) = \kappa^e(0)(1 - \omega_c^2 \tau^2 / 4)$ (*32*). Clearly, the renormalized factor is 4 times smaller than that for the topological materials with the linear Dirac band due to the different relationship between energy and effective mass for parabolic and linear band. Additionally, the parabolic band has a much smaller Fermi-velocity compared to the linear Dirac band, so the reduction of electric thermal conductivity by the perpendicular magnetic field is very weak in the conventional materials. Secondly, the lattice part is usually dominant in the thermal conductivity at high temperature in conventional thermoelectric materials. It is well known that magnetic field can tune the electronic thermal conductivity, while has negligible effect on the lattice thermal conductivity. As discussed above, in order to greatly enhance *ZT* by magnetic field, the materials should have two characteristics. One is the linear Dirac band with large Fermi-velocity which is the character of topological materials, and the other is that the thermal conductivity is dominated by the electric thermal conductivity at high temperature. Fortunately, the 3D topological Dirac semimetal $Cd_3As_2$ owns the two characteristics, and our work shows a large thermomagnetic effect in $Cd_3As_2$ crystal.

## 4. Summary

We reported magneto-thermoelectric figure of merit in three-dimensional Dirac semimetal $Cd_3As_2$ crystals with [112] and [100] orientation, respectively. It is found that the magnetic field dramatically reduces electric thermal conductivity and enhances the thermoelectric power, consequently leads to a dramatic enhancement of *ZT*. The huge enhancement of *ZT* by magnetic field could be related to the linear Dirac band in Cd$_3$As$_2$. Our work provides a new way to effectively enhance *ZT* by magnetic field for the topological materials with linear energy dispersion, and hopefully exploit the thermoelectric performance of topological materials.

**Conflict of interest**

The authors declare that they have no conflict of interest.

**Acknowledgements**

This work is supported by the National Key R&D Program of the MOST of China (Grant Nos. 2017YFA0303001, 2016YFA0300201 and 2017YFA0204904), the Nature Science Foun-dation of China (Grant Nos.11534010, 11774325 and 21603210), the Key Research Program of Frontier Sciences CAS (Grant No. QYZDY-SSW-SLH021), Hefei Science Center CAS (2016HSC-IU001), and the Fundamental Research Funds for the Central Universities. We also thank Supercomputing Center at USTC for providing the computing resources. This work was partially performed on the Superconducting Magnet and PPMS-16T Facilities, High Magnetic Field Laboratory of CAS.

**Appendix A. Supplementary data**

Supplementary data associated with this article can be found, in the online version.

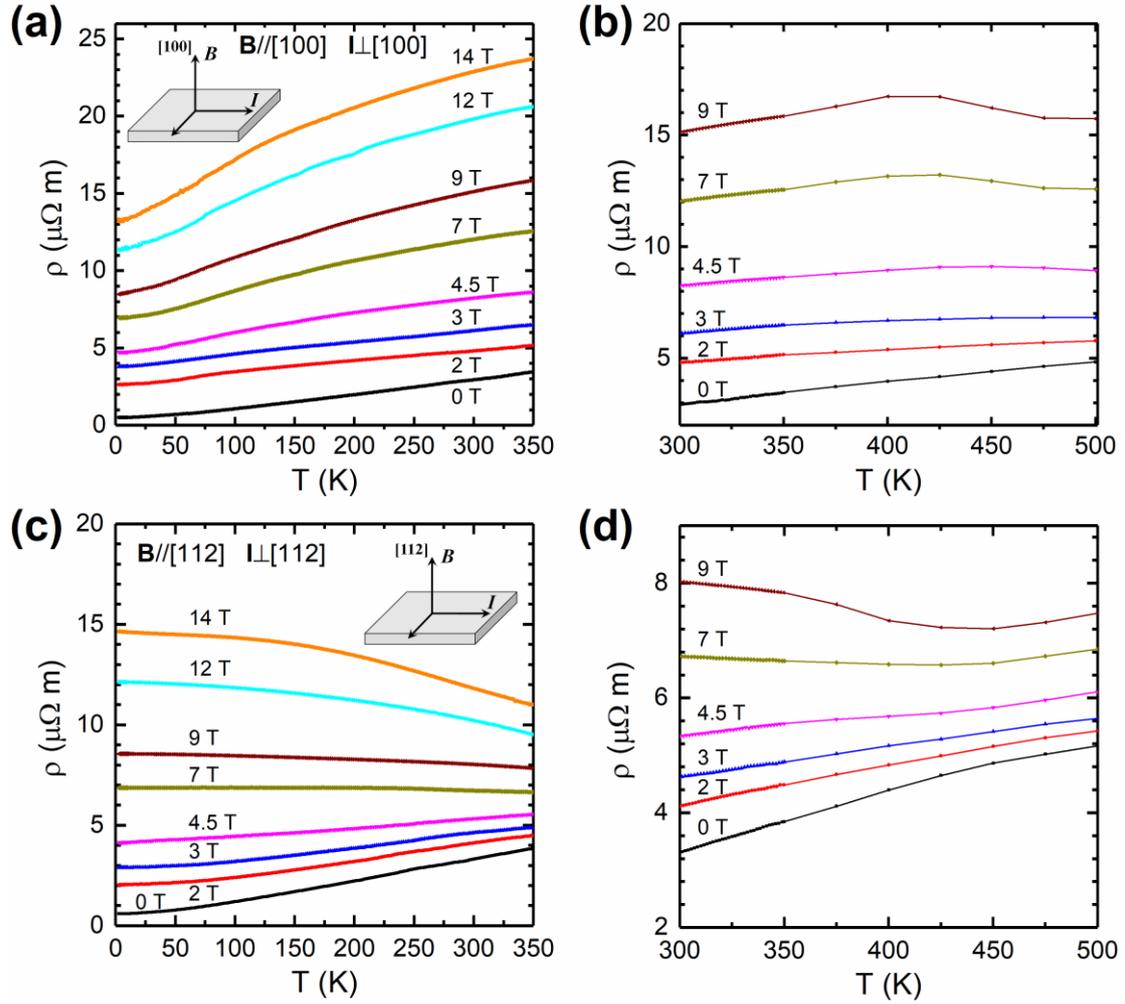

**Fig. 1** (color online) Temperature dependence of resistivity under various magnetic fields for the $Cd_3As_2$ single crystals with [100] or [112] orientation. (a) and (b): The external magnetic field is applied along [100] direction, and electric current is perpendicular to the [100] direction. (c) and (d): The external magnetic field is applied along [112] direction, and electric current is perpendicular to [112] direction.

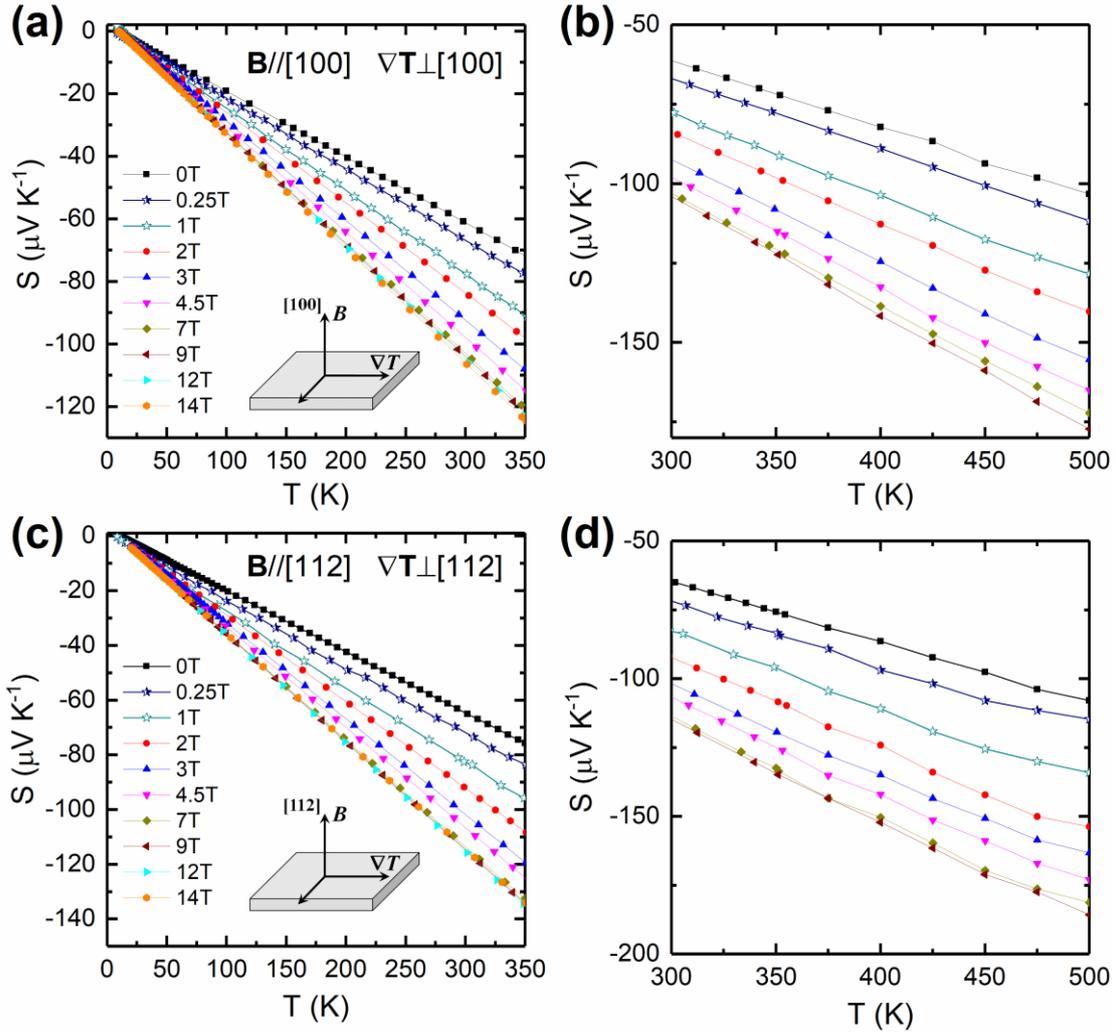

**Fig. 2** (color online) Temperature dependence of thermopower under various magnetic fields for the $Cd_3As_2$ single crystals with [100] or [112] orientation. (a) and (b): The external magnetic field is applied along [100] direction, and temperature gradient is perpendicular to the [100] direction. (c) and (d): The external magnetic field is applied along [112] direction, and temperature gradient is perpendicular to [112] direction.

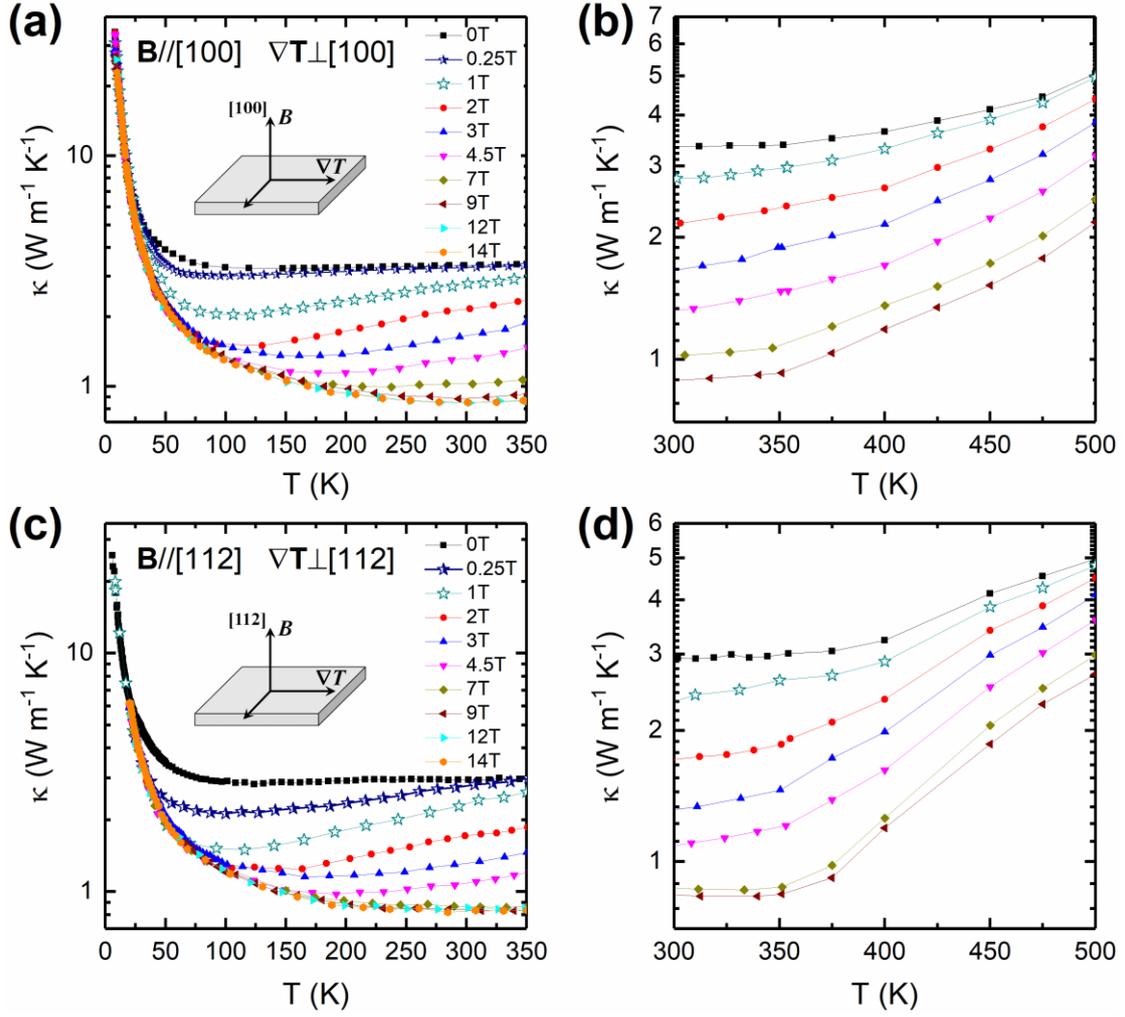

**Fig. 3** (color online) Thermal conductivity as a function of temperature under different magnetic fields for $Cd_3As_2$ single crystals with [100] or [112] orientation. (a) and (b): The external magnetic field is applied along [100] direction, and temperature gradient is perpendicular to the [100] direction. (c) and (d): The external magnetic field is applied along [112] direction, and temperature gradient is perpendicular to [112] direction.

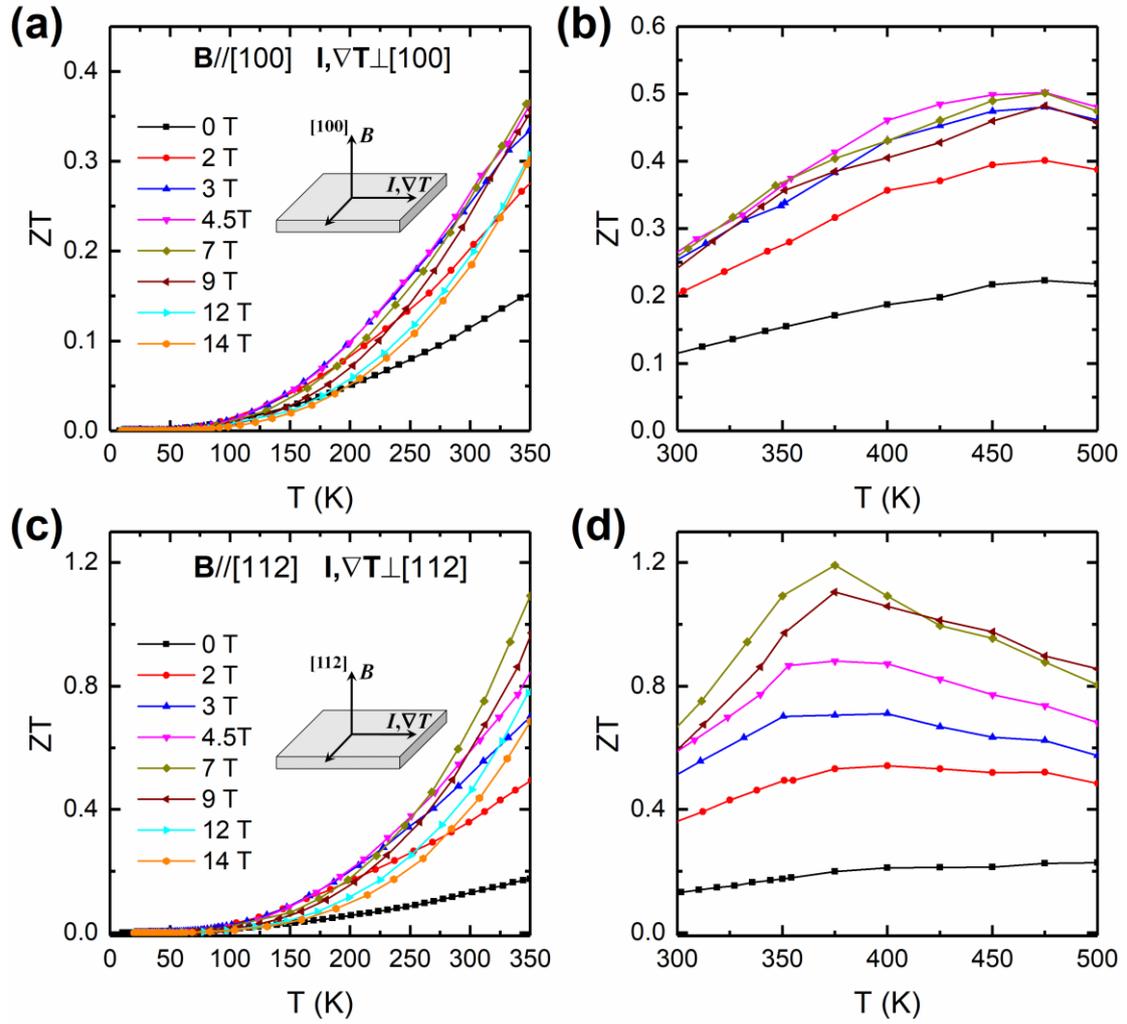

**Fig. 4** (color online) The thermoelectric figure of merit as a function of temperature under different magnetic fields for $Cd_3As_2$ single crystals with [100] or [112] orientation. (a) and (b): The external magnetic field is applied along [100] direction, and electric current or temperature gradient is perpendicular to the [100] direction. (c) and (d): The external magnetic field is applied along [112] direction, and electric current or temperature gradient is perpendicular to [112] direction.